\documentclass[twocolumn]{aastex631}

\usepackage[T1]{fontenc}
\usepackage[utf8]{inputenc}
\usepackage{float}
\newcommand{\bpic}[1]{$\beta$ Pic}
\newcommand{\apic}[1]{$\alpha$ Pic}

\usepackage{graphicx}
\usepackage{amsmath}

\shorttitle{Temporal Evolution of the Beta Pic Debris Disk}
\shortauthors{Avsar et al.}

\begin{document}

%\title{26 Years of High Precision HST/STIS Imaging of the Beta Pictoris Debris Disk: A Search for Collisions and Planet-Disk Interactions}

\title{A Search for Collisions and Planet-Disk Interactions in the Beta Pictoris Disk with 26 Years of High Precision HST/STIS Imaging}

\author[0000-0001-7801-7425]{Arin M. Avsar}
\affiliation{The Lunar and Planetary Laboratory, University of Arizona, USA}
\author[0000-0002-4309-6343]{Kevin Wagner}
\affiliation{Steward Observatory, University of Arizona, USA}
\author[0000-0003-3714-5855]{Dániel Apai}
\affiliation{Steward Observatory, University of Arizona, USA}
\affiliation{The Lunar and Planetary Laboratory, University of Arizona, USA}

\author{Christopher C. Stark}
\affiliation{NASA Goddard Space Flight Center, Exoplanets and Stellar Astrophysics Laboratory, Code 667, Greenbelt, MD 20771, USA}

\author[0000-0001-9064-5598]{Mark C. Wyatt}
\affiliation{Institute of Astronomy, University of Cambridge, Madingley Road, Cambridge CB3 0HA, UK}

\begin{abstract}

$\beta$ Pictoris (\bpic{})’s well-studied debris disk and two known giant planets, in combination with the stability of HST/STIS (and now also JWST), offers a unique opportunity to test planet-disk interaction models and to observe recent planetesimal collisions. We present HST/STIS coronagraphic imaging from two new epochs of data taken between 2021 and 2023, complementing earlier data taken in 1997 and 2012. This dataset enables the longest baseline and highest precision temporal comparison of any debris disk to date, with sensitivity to temporal surface brightness variations of sub-percentage levels in the midplane of the disk. While no localized surface brightness changes are detected, which would be indicative of a recent planetesimal collision, there is a tentative brightening of the SE side of the disk over the past decade. We link the constraints on surface brightness variations to dynamical models of the planetary system's evolution and to the collisional history of planetesimals. Using a coupled collisional model and injection/recovery framework, we estimate sensitvity to expanding collisional debris down to a Ceres-mass per progenitor in the most sensitive regions of the disk midplane. These results demonstrate the capabilities of long-baseline, temporal studies with HST (and also soon with JWST) for constraining the physical processes occurring within debris disks.
\end{abstract}


\section{Introduction}

Debris disks are found during the late stages of planet formation, where most of the gas in the protoplanetary disk has dissipated and micron-sized dust grains to kilometer-sized debris are produced as a result of collisional cascades of planetesimals. Spatially resolved debris disks offer insights into the dynamical history of young planetary systems through linking their surface brightness distributions to underlying morphologies like rings, gaps, asymmetries, etc. (see reviews by \citealt{2008ARA&A..46..339W,Hughes_Review2018, Wyatt2020}). 

Although it is generally accepted that dust in debris disks are sustained by the continual collision of planetesimals, but only few such initial collisional remnants have been directly observed or identified \citep{extreme_disks_spitzer,2020PNAS..117.9712G}. Specifically, the transient brightness signatures produced in the immediate aftermath of a large planetesimal collision have remained elusive, since such signatures last less than one orbital period. This transient brightness signature should be distinguished from the longer term variations larger collisional debris (> 1-2 $\mu$m) can have on on disk morphologies and overall surface brightness asymmetries over a period of hundreds to millions of years \citep[e.g.][]{2014MNRAS.440.3757J,2015A&A...573A..39K,2023ApJ...948..102J}. The reason why the initial transient brightness signatures are difficult to find is in part due to the fact that such signals do not last more than a couple decades in scattered light since stellar radiation pressure disperses or expels the smaller dust grains from the system. With decades-long instrumental stability, signals of planetesimal collisions might be detectable, especially in young planetary systems where such collisions occur more frequently \citep{Young_system_collisions}. For this reason, temporal monitoring of resolved debris disks can offer us unique insight and opportunities to observe such events and constrain theoretical models by identifying and observing the morphological and spectroscopic evolution of collisions over time.

The Hubble Space Telescope's Space Telescope Imaging Spectrograph (hereby referred to as HST/STIS) was commissioned in 1997 \citep{STIS_Design} and now presents an opportunity to study changing dynamical features over a period of decades in nearby debris disks \citep{2018ARA&A..56..541H}. These can then be followed up with JWST $-$ e.g., in spectroscopic modes to study the underlying grain distribution and thermal emission from giant planets. For example, fast-moving dust features around the M-dwarf AU Mic have been seen to traverse dozens of au over a timespan of less than a decade \citep{2015Natur.526..230B, 2018A&A...614A..52B, 2020ApJ...889L..21G}, and a multi-epoch analysis of STIS data from the debris ring around Fomalhaut has found evidence of a recent large planetesimal collision \citep{2008Sci...322.1345K, 2020PNAS..117.9712G}. Recent JWST observations have not found large separation giant planets down to 0.1 M$_J$ for the case of AU Mic \citep{AUMIC_JWST} and has found evidence for a potential candidate with a mass of 1 M$_J$ for the case of Fomalhaut \citep{Fomalhaut_JWST}. Currently, $\beta$ Pictoris ($\beta$ Pic) is the only debris disk-hosting system with a confirmed large separation giant planet \citep{Beta-Pic-b-discovery,Beta-Pic-b-confirm} whose influence on the debris disk can be observed and studied over time \citep{2015ApJ...800..136A}.

$\beta$ Pic is an A6V star located at a distance of 19.83 parsecs with an estimated age of $18.5^{+2.0}_{-2.4}$ Myr \citep{gaia_beta_pic, 2020A&A...642A.179M}. The system is known to host two super-Jupiter exoplanets. \bpic{} b is the outer, directly-imaged companion with a semi-major axis of 10.26 $\pm$ 0.10 au \citep{Beta-Pic-b-discovery,Beta-Pic-b-confirm,2021AJ....161..179B}. Meanwhile, the recently discovered \bpic{} c is at a separation of $2.72 \pm 0.02$ au and was detected through interferometric and radial velocity observations \citep{Beta-Pic-c-discovery}. Additionally, \bpic{} hosts a bright debris disk, $F_{disk}/F_{\star} \approx 2.5 * 10^{-3}$ \citep{2000prpl.conf..639L}, that extends over 1000 au \citep{1984Sci...226.1421S,2001MNRAS.323..402L, 2021A&A...646A.132J}. \bpic{} b's relatively short orbital period ($\sim$ 20 years), coupled with multi-epoch imaging of the disk, provides a unique opportunity to study the influence of a giant planet on a debris disk over dynamical timescales (i.e., similar to the planet's orbital period) \citep{2018ARA&A..56..541H}.

Being one of the most well-studied debris and planetary systems, a variety of features have been identified within the \bpic{} disk. Mid-IR observations of the disk have found a large dust clump on the western side of the disk \citep{Telesco_MIR}. Further sub-mm observations by ALMA have detected CO patch co-located with the mid-IR observations, possibly created by the aftermath of a major planetesimal collision \citep{2014Sci...343.1490D} or a collisional avalanche of small particles \citep{1997AREPS..25..175A, 2007A&A...461..537G}. Whether the dust clump is moving or stationary is still uncertain, with each scenario having different implications for the origin of the clump \citep{Skaf_MIR,Han_MIR}. A moving clump could imply a vortex of gas trapping the dust by an unseen planet, while a stationary clump implies the occurrence of a past large collision. Furthermore, recent results from JWST observations of the system using the MIRI instrument \citep{MIRI-Intro} have identified a ``cat's tail''-like feature in the disk \citep{JWST_BPic}. The feature is thought to be the remnant of two planetesimals colliding in the disk, roughly the size of a small moon of Saturn or Jupiter. 

\bpic{} b is thought to have significant influence on the surrounding debris disk, with the location and mass of the planet being predicted through the notable 3$^\circ$ warp of the main planetesimal belt \citep{beta_pic_b_prediction}. Furthermore, modeling has shown that the inclination of planetesimals under the influence of the \bpic{} b will oscillate between 0 and 2$i_p$, where this $i_p$ is the inclination of \bpic{} b with respect to the disk midplane \citep{2011ApJ...743L..17D}. The oscillation of planetesimal inclinations and eccentricities creates an environment where planetesimal collisions are more likely to be destructive \citep{stirring_collisions}. 

The goal of this study will be to use long baseline STIS observations of the \bpic{} system to identify changing disk structures potentially driven by the orbit of \bpic{} b. Furthermore, we search for and model debris clouds created from the immediate aftermath of a stochastic giant planetesimal collision, and address our sensitivity to dust clouds compromised of mainly sub-micron sized grains created by future such collisions in the \bpic{} disk with HST/STIS. This study will not address the effect that such collisions will have on disk morphologies on the timescales of hundreds to millions of years \citep{2014MNRAS.440.3757J, 2023ApJ...948..102J} or larger radii debris remnants (> 1-2 $\mu$m) \citep{2015A&A...573A..39K, 2015ApJ...810..136G}. 
~ %moving header to next column

\section{Observations}

In this paper we present the reduction of three epochs of STIS observations of \bpic{}, with one archival dataset (2012) and two new epochs (2021, 2023). All epochs were taken in the \texttt{50CORON} (coronagraphic) imaging mode and with identical occulter configurations. This mode has a broad bandpass ranging from 200 to 1050 nm. The STIS CCD has dimensions of 1024x1024 pixels and a plate scale of 0$\farcs$05077 per pixel \citep{STIS-DG}.

\subsection{2012-2023 STIS Observations}

The 2012 data were acquired as part of the program GO-12551 (PI: Apai). The observations were taken in three orbits, with the first and third orbits imaging $\beta$ Pic and the second orbit imaging the PSF reference star, $\alpha$ Pic. $\alpha$ Pic was chosen to be the PSF reference star due to its similar color and apparent magnitude to \bpic{} (V = 3.30, $\Delta$(B-V) w.r.t. \bpic{} = -0.01), as well as its small angular distance from the target. Spacecraft rolls were performed during each orbit in order to image all sections of the disk that would have otherwise been blocked or obscured by the occulting wedges or diffraction spikes. During each orbit, \texttt{WEDGEA0.6} (0$\farcs$6 width), \texttt{WEDGEA1.0} (1$\farcs$0 width), and \texttt{WEDGEB1.0} (1$\farcs$0 width) occulters were used to take short, medium, and long exposures. The short exposures used the \texttt{WEDGEA0.6} and \texttt{WEDGEB1.0} occulters to image the inner disk and the stellar PSF for both the target and reference star. The long exposures used \texttt{WEDGEA1.0} and \texttt{WEDGEB1.0} occulters to image the outer disk (past 2\farcs), while saturating the inner disk. 

The 2021 and 2023 data were acquired as a part of program GO-16788 (PI: Wagner). The data from both epochs were acquired in identical occulter and exposure time configurations to the 2012 epoch, except for the last \texttt{WEDGEB1.0} exposures for each orbit being unable to execute for the 2023 data due to scheduling conflicts.

\begin{deluxetable*}{cccccc}
\tabletypesize{\footnotesize}
\tablecolumns{6}
\tablewidth{0pt}
\tablecaption{ Summary of the three epochs of STIS observations of \bpic{} and \apic{}}
\tablehead{
\colhead{Program}  & \colhead{Date} &  \colhead{Visit \#} & \colhead{Target} & \colhead{Int. Time [s]} & \colhead{Occulter}\\}
\startdata
12551 & 03/06/2012 & 1 & \bpic{} &$11\times1.2$ & WEDGEA0.6 \\ 
12551 & 03/06/2012 & 1 & \bpic{} &$4\times60.0$,$16\times3.0$ & WEDGEA1.0 \\ 
12551 & 03/06/2012 & 1 & \bpic{} & $4\times 60.0$, $16\times3.0$ & WEDGEB1.0 \\ 
12551 & 03/06/2012 & 1 & \bpic{} & $11\times1.2$ & WEDGEB0.6 \\
12551 & 03/06/2012 & 2 & \apic{} & $11\times0.7$ & WEDGEA0.6 \\ 
12551 & 03/06/2012 & 2 & \apic{} & $4\times36.0$ ,$16\times1.9$ & WEDGEA1.0 \\
12551 & 03/06/2012 & 2 & \apic{} & $4\times36.0$, $16\times1.9$ & WEDGEB1.0 \\ 
12551 & 03/06/2012 & 2 & \apic{} & $17\times0.7$ & WEDGEB0.6 \\ 
12551 & 03/06/2012 & 3 & \bpic{} &$11\times1.2$ & WEDGEA0.6 \\ 
12551 & 03/06/2012 & 3 & \bpic{} &$4\times60.0$,$16\times3.0$ & WEDGEA1.0 \\ 
12551 & 03/06/2012 & 3 & \bpic{} & $4\times 60.0$, $16\times3.0$ & WEDGEB1.0 \\ 
12551 & 03/06/2012 & 3 & \bpic{} & $11\times1.2$ & WEDGEB0.6 \\
\hline
16788 & 02/28/2021 & 1 & \bpic{} & $8\times1.2$ & WEDGEA0.6 \\
16788 & 02/28/2021 & 1 & \bpic{} & $4\times60.0$, $16\times3.0$ & WEDGEA1.0 \\ 
16788 & 02/28/2021 & 1 & \bpic{} & $4\times60.0$, $16\times3.0$ & WEDGEB1.0 \\ 
16788 & 02/28/2021 & 1 & \bpic{} & $9\times1.2$ & WEDGEB0.6 \\ 
16788 & 02/28/2021 & 2 & \apic{} & $11\times0.7$ & WEDGEA0.6 \\ 
16788 & 02/28/2021 & 2 & \apic{} & $4\times36.0$, $16\times1.9$ & WEDGEA1.0 \\ 
16788 & 02/28/2021 & 2 & \apic{} & $4\times36.0$, $15\times1.9$ & WEDGEB1.0 \\ 
16788 & 02/28/2021 & 2 & \apic{} & $14\times0.7$ & WEDGEB0.6 \\
16788 & 02/28/2021 & 3 & \bpic{} & $9\times1.2$ & WEDGEA0.6 \\ 
16788 & 02/28/2021 & 3 & \bpic{} & $4\times60.0$, $16\times3.0$ & WEDGEA1.0 \\ 
16788 & 02/28/2021 & 3 & \bpic{} & $4\times60.0$, $16\times3.0$ & WEDGEB1.0 \\ 
16788 & 02/28/2021 & 3 & \bpic{} & $8\times1.2$ & WEDGEB0.6 \\ 
\hline
16788 & 03/06/2023 & 1 & \bpic{} & $8\times1.2$ & WEDGEA0.6 \\ 
16788 & 03/06/2023 & 1 & \bpic{} & $4\times60.0$, $16\times3.0$ & WEDGEA1.0 \\ 
16788 & 03/06/2023 & 1 & \bpic{} & $4\times60.0$, $16\times3.0$ & WEDGEB1.0 \\ 
16788 & 03/06/2023 & 1 & \bpic{} & $9\times1.2$ & WEDGEB0.6 \\ 
16788 & 03/06/2023 & 2 & \apic{} & $11\times0.7$ & WEDGEA0.6 \\ 
16788 & 03/06/2023 & 2 & \apic{} & $4\times36.0$, $16\times3.0$ & WEDGEA1.0 \\ 
16788 & 03/06/2023 & 2 & \apic{} & $4\times36.0$, $16\times3.0$ & WEDGEB1.0 \\ 
16788 & 03/06/2023 & 2 & \apic{} & $14\times0.7$ & WEDGEB0.6 \\ 
16788 & 03/06/2023 & 3 & \bpic{} & $9\times1.2$ & WEDGEA0.6 \\ 
16788 & 03/06/2023 & 3 & \bpic{} & $4\times60.0$, $16\times3.0$ & WEDGEA1.0 \\ 
16788 & 03/06/2023 & 3 & \bpic{} & $4\times60.0$, $16\times3.0$ & WEDGEB1.0 \\ 
16788 & 03/06/2023 & 3 & \bpic{} & $8\times1.2$ & WEDGEB0.6 \\ 
\enddata
\vspace{-0.8cm}
\end{deluxetable*}

%annotate diffraction spike figure

\section{STIS Data Reduction}

For our data reduction, we downloaded the the cosmic ray rejected and flat fielded files, {\tt\string crj}, from the Mikulski Archive for Space Telescopes (MAST) Portal. We first conducted a background subtraction of each raw image by subtracting the mean pixel value in a 30 pixel radius aperture, approximately 200 pixels away from the location of the coronagraphic wedge. Then, each raw image was converted from counts to counts per second by dividing each image by its respective exposure time. 

We calculate the the location of the star behind the occulting wedge, using the package {\tt\string centerRadon} for both target and PSF reference images \citep{2017SPIE10400E..21R}. Using the calculated star positions, each image was shifted to a common center using {\tt\string scipy.ndiamge.interpolation}. 

We follow similar PSF subtraction steps to those described in \citet{2015ApJ...800..136A}. We conduct a grid search of the PSF reference images to best subtract the starlight from \bpic{}, by minimizing the diffraction spike residuals after the subtraction of the reference star. This is done by varying three parameters in the grid search: $x$-position, $y$-position, and intensity. By varying the the $x$ and $y$ parameters, we make sure that the PSF reference image, specifically the star behind the occulter, is best aligned with the target image for subtraction. The intensity parameter (ranging from 0.0 to 1.0 in our case) ensures that the flux from \apic{} is not over-subtracting or under-subtracting the starlight from \bpic{}, since the reference star can have an apparent magnitude not equal to that of the target star. Before the grid search, we mask out the central region containing significant disk light. This is done to ensure the diffraction spike residuals that we are minimizing comes only from the star and is not contaminated by the disk scattered light. Without masking the inner disk region, over-subtraction of disk light occurs. The size of the mask differed for different occulting wedge and integration times but ranged between 30-100 pixels in radius. Finally, we minimize the diffraction spike residuals, defined as the sum of squared residual pixel fluxes in the diffraction spikes. We iterate through all three parameters and choose the combination that minimizes the sum of the diffraction spike residuals. 

\begin{figure}[!ht]
    \center
    \includegraphics[width=0.45\textwidth]{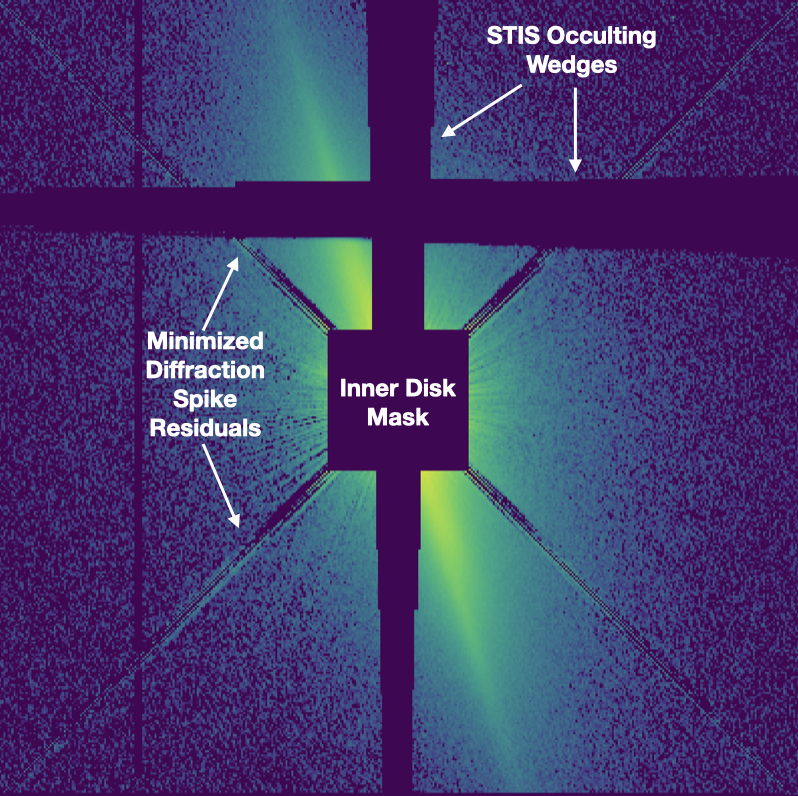}
    \caption{An example of PSF subtraction regions for our automated pipeline. Before the grid search, the inner 30x30 pixels of the disk is masked so that we are only minimizing diffraction spike light that is not contaminated by disk light. Our grid search, described in Section 3, iterates through all parameters until it finds the combination that minimizes the square sum of the diffraction spike residuals pictured.}
    \label{fig:spike_image}
\end{figure}

\begin{figure*}%[!ht]
    \center
    \includegraphics[width=\textwidth]{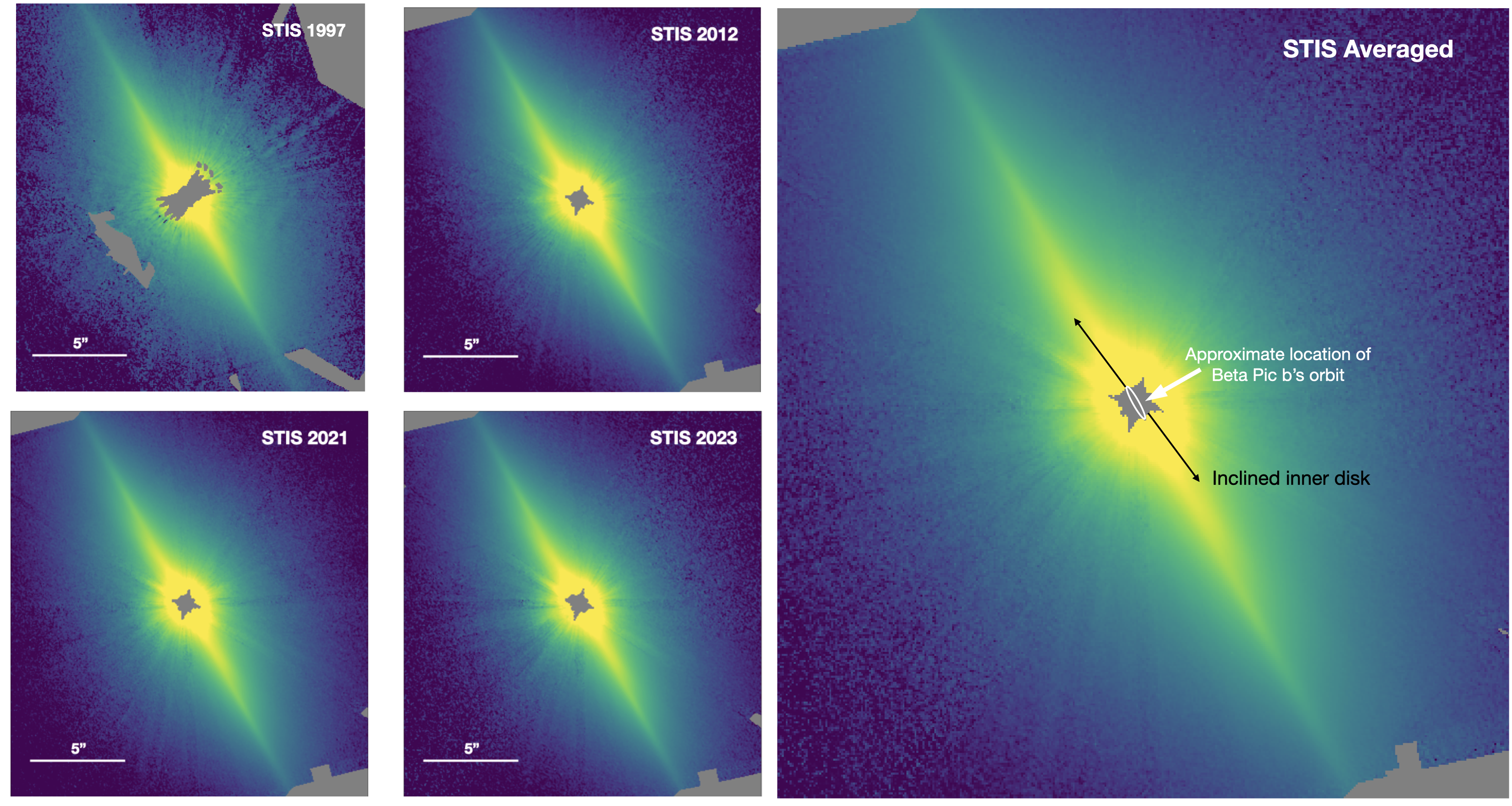}
    \caption{Visual comparison of four epochs of the \bpic{} disk. The top left panels show two archival images. The 1997 image is reproduced from \cite{2015ApJ...800..136A}, while we re-reduce the 2012 dataset. The bottom left panels show two new STIS coronagraphic images of the \bpic{} disk (2021, 2023). The right panel shows the mean combined image of the 2012-2023 datasets, along with the orbit of \bpic{} b relative to the disk and the orientation of the inclined inner disk \citep[]{beta_pic_acs,2015ApJ...800..136A}. The images are shown on a logarithmic stretch ranging from 0.01-100 counts per second. All images are rotated to the common north up, east left orientation. }
    \label{fig:disk_gallery}
\end{figure*}

All PSF-subtracted target images are derotated into the "north up" and "east left" orientation before being mean combined. The final images of the outlined reduction steps are shown in \ref{fig:disk_gallery}. 

Finally, the uncertainty maps are generated by taking the standard error of the mean for each pixel across all combined images. How error is propagated in the disk image analyses can be found in the Appendix. 

~

\section{Observational Results}

Although the 1997 data is shown in Figure \ref{fig:disk_gallery}, we do not directly incorporate the 1997 data in our analysis due to the different imaging orientation and lower data quality compared to the 2012-2023 datasets. For temporal analysis incorporating the 1997 data, see \cite{2015ApJ...800..136A}. However, we will compare our results to those found in the analysis by \cite{2015ApJ...800..136A}.

%\begin{figure*}[!ht]
%    \center
%    \includegraphics[width=\textwidth]{Figures/ratio_and_r2.png}
%    \caption{A comparison of the three epochs of data. The top three panels compare the radially normalized images of the \bpic{} disk. This was done by multiplying the disk images by $r^2$, where r is the stellocentric distance at each point in the image. The bottom two panels compare the ratio maps, which was made by dividing the un-normalized images from one another (2021/2012 and 2023/2012). A Gaussian filter was applied to both images to enhance brightening features in order to search for planetesimal collision candidates. One candidate was found and is marked by the white circle in both images.} 
%\end{figure*}

\begin{figure*}[!ht]
    \center
    \includegraphics[width=\textwidth]{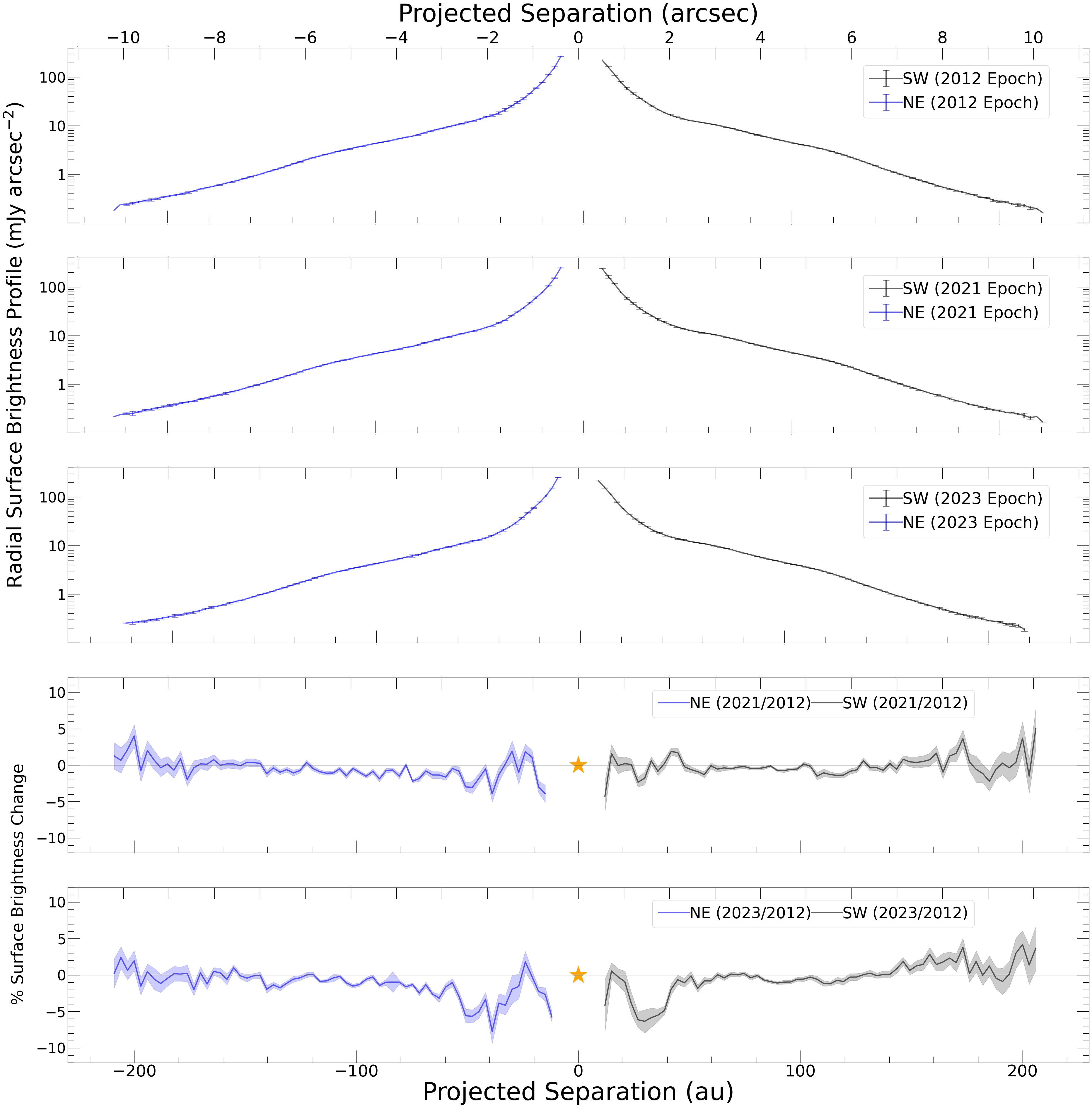}
    \caption{Midplane radial surface brightness measurement and epoch comparison. The top three panels show individual radial surface brightness measurements along the disk midplane of each epoch in units of mJy arcsec$^{-2}$. The bottom two panels show the percent change in midplane radial surface brightness profiles between epochs (2021/2012 and 2023/2012), with the shaded region around the lines indicating the 1$\sigma$ uncertainty. Aperture photometry along the disk midplane show observed variation to be consistent with noise in the data to 3$\sigma$.}
    \label{fig:SB_map}
\end{figure*}

\begin{figure*}[!ht]
    \center
    \includegraphics[width=\textwidth]{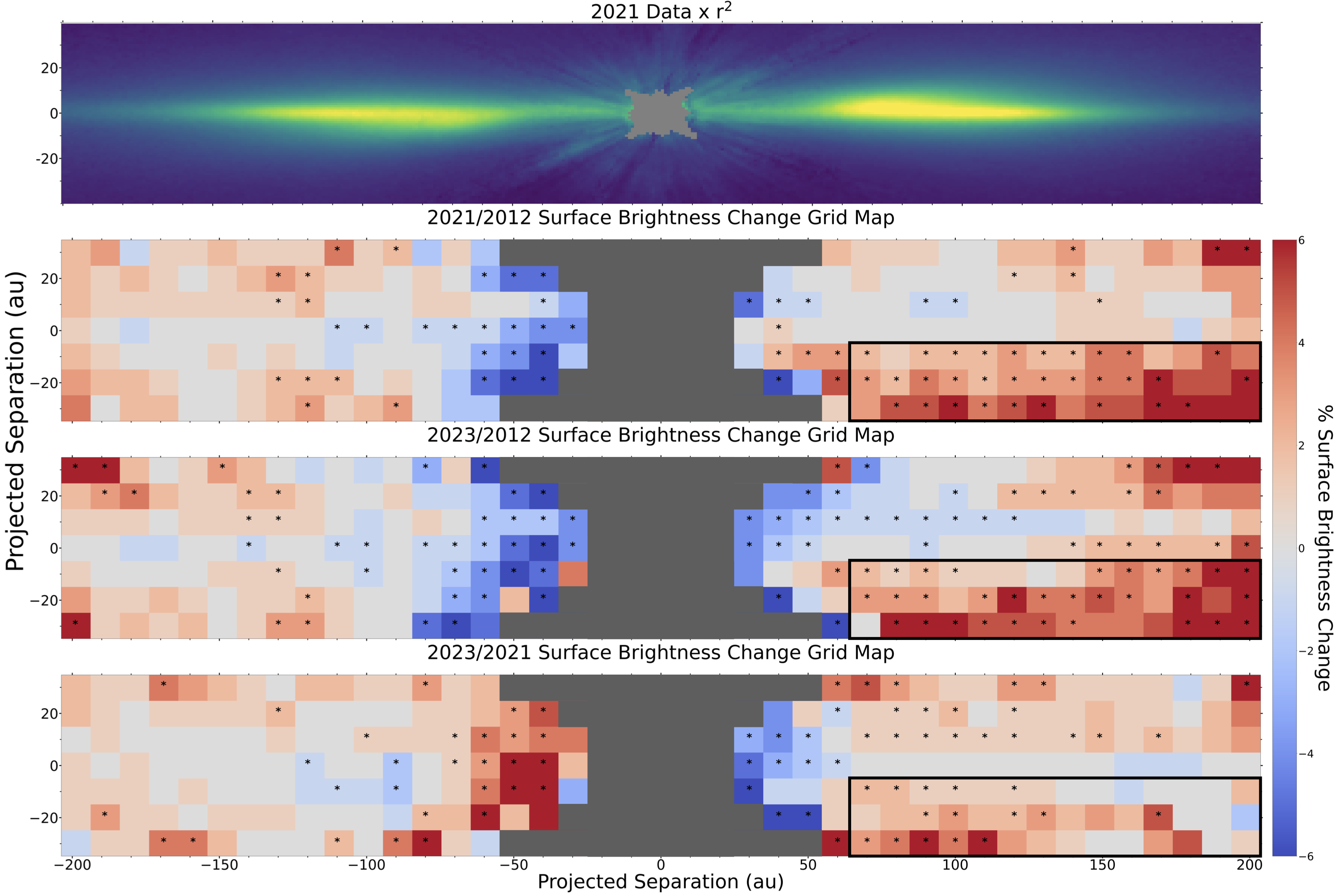}
    \caption{Ratio grid maps showing the percent surface brightness change between epochs and associated 1$\sigma$ uncertainties in each 10 au x 10 au grid cell. The astero indicate cells where the SNR of the surface brightness change is $\ge 5$. The area within the grey region is blocked out due to the disk surface brightness is dominated by PSF-subtraction residuals. The top image should be used as a reference to where in the disk each grid cell is probing. We can see in the 2023/2012 and 2021/2012 ratio grid maps a significant, consistent brightening of the disk below the western midplane. This region is shown in the black boxed area. }
    \label{fig:ratio_grid}
\end{figure*}

\subsection{HST/STIS Images of the \bpic{} Disk and Data Signal-To-Noise By Disk Region}

The instrument imaging orientations shown in Table 1 allow us to reach an inner working angle of $\sim$ 10 au with an field-of-view out to 220 au. The inner working angle (IWA)is defined as the inner edge of our occulting wedge. Our IWA is located at the edge of \bpic{} b's orbit, which is not detected in our images due to the scattered light from the disk dominating the signal. 

We find little to no visual structural changes in the disk between epochs. We resolve dust density asymmetries between the northeast and southwest sides of the planetesimal belt reported in MIR and sub-mm observations \citep[][]{Skaf_MIR,2014Sci...343.1490D}{}{}, which are prominently seen between 2$\farcs$- 6$\farcs$ in the top panel of Figure \ref{fig:ratio_grid}. 

The highest SNR region of the disk is located between 50-150 au (or 2.5$\farcs$-7.5$\farcs$). Past 150 au, the low dust density coupled with the $r^{-2}$ falloff of signal in scattered light observations causes the SNR to decrease as well. Meanwhile within 50 au, the disk is brighter, but higher noise levels contribute to the decrease in SNR observed. More specifically, increased photon noise due to a brighter disk at smaller separations coupled with residual speckles from PSF subtraction increase the overall noise interior of 50 au compared to larger separations.

\subsection{Search For Planetesimal Collision Candidates}

From the disk images, we searched for surface brightness variations in the disk between the three epochs that could be the result of a planetesimal collision in disk. 

The identification of potential planetesimal collision signals was done by first dividing the 2021 and 2023 images by the 2012 image to create two ratio maps comparing the surface brightness change in the disk over time. First, we searched for signals where a brightening in the disk is seen in both 2021/2012 and 2023/2012 ratio maps. A brightening in the disk over time could indicate a recent planetesimal collision.

Using both visual inspection and midplane radial surface brightness profiles, no point sources or resolved candidates were identified with an SNR of greater than 2.1 and consistent location in the disk between epochs, implying the origin of the signals are most likely random noise.  

\begin{figure*}[!ht]
    \centering
    \includegraphics[width=\textwidth]{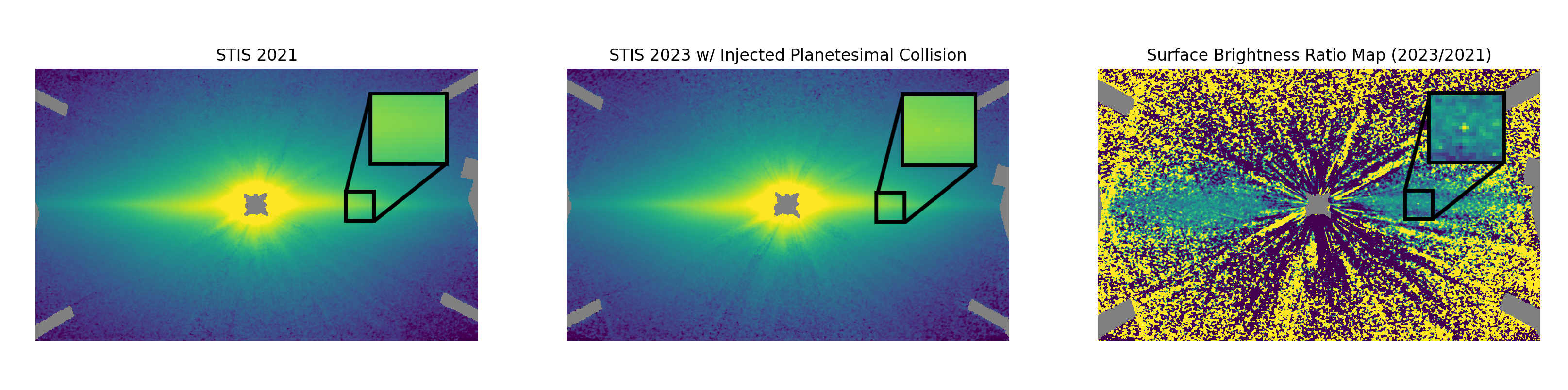}
    \caption{Simulation of two 15 Ceres mass planetesimals colliding at 100 au in the \bpic{} disk. The collision was simulated as if it occurred right after our 2021 epoch of data was taken and is then injected in the 2023 epoch of data. The collision is very difficult to see in the 2023 epoch alone due to the high surface brightness of the disk. However, taking a ratio between the two epochs allows the excess surface brightness from the collision to stand out as a clear detection. No such signals are seen in any of the ratio maps, allowing us to rule out any new major collisions in the disk. }
    \label{fig:injection}
\end{figure*}

\subsection{Temporal Changes Along The Disk Midplane}

A major goal of this study is to measure midplane surface brightness variations between epochs. In order to measure surface brightness variations of the disk along the midplane, one pixel radius apertures were placed along the midplane of the disk. One pixel radius apertures were chosen due it being the size of a STIS resolution element. The placement of the aperture was determined by taking the brightest point of a vertical cut of the disk midplane in the 2012 epoch. The 2012 aperture locations were also used to determine the surface brightness profiles of the 2021 and 2023 epochs in order to accurately compare surface brightness changes between epochs. We then determined the midplane locations of the 2021 and 2023 epochs, finding that the location of the midplane from 2012 to 2021 and 2023 has not changed by more than a tenth of a pixel in all locations of the disk. This ensures that the one pixel radius aperture accurately samples the midplane of all disk images. Finally, we take the ratio between radial surface brightness measurements from each epoch in order to find the variation over time. The result of the midplane measurements can found in Figure \ref{fig:SB_map}. 

Overall, we find very little statistically significant variation in all regions of the disk midplane between 2012 and 2023. In the inner disk (< 50 au), we find surface brightness changes on the order of 4\% for the 2021/2012 comparison and 7\% for the 2023/2012. In the region hosting the main planetesimal belt of \bpic{} (50-150 au), we constrain surface brightness variations of $\le$ 1.5\% in all epoch comparisons. Finally in the outer disk ($\ge$ 150 au) of \bpic{}, we constrain surface brightness changes on the order of $\le$ 5\% between all epoch comparisons.

% exhibiting an average surface brightness ratio value of 0.943 $\pm$ 0.019.  and has an average surface brightness ratio value 0.945 $\pm$ 0.022. NE
\subsection{Quantifying Temporal Variations Throughout The \bpic{} Disk}

Although the midplane ratio measurements in Figure \ref{fig:SB_map} provide a general idea of how the surface brightness in the \bpic{} disk has changed in the past eleven years, they do not paint a full picture of how the entire disk has changed. To provide disk surface brightness variations for the entire disk, we created a grid ratio map. We first divided the ratio maps into 10x10 au grid cells and then computed the mean percent surface brightness change and associated 1$\sigma$ uncertainty in each grid cell, which is also printed in each cell. The final grid maps are found in Figure \ref{fig:ratio_grid}.

Unlike in the radial surface brightness profiles, the grid maps allow us to quantify changes outside the disk midplane. Specifically, they show a surface brightness increase below of the western midplane from 80-200 au in both the 2021/2012 and 2023/2012 grid maps. Integrating over the whole region, we measure a mean surface brightness increase of 4.97\% $\pm$ 0.88\% in the 2021/2012 grid map and 4.52\% $\pm$ 0.89\% in the 2023/2012 grid map, making the SNR of both measurements > 5. This feature is not recovered in the 2023/2021 data or in other quadrants of the disk in any epochal comparison, suggesting a surface brightness change in the 2012 epoch. At this stage, we cannot rule out the brightness variations being artifacts of a variable stellar halo in either the science or reference images in the 2012 data since the surface brightness in this part of the disk ranges from 0.2-0.5 mJy arcsec$^{-2}$. 

If the brightness differences seen in our data are of astrophysical nature, a deeper exploration of their origins will be compelling. Here, we will explore the dust created at the immediate aftermath of massive planetesimal collisions as one possible mechanism. 

\section{Modeling Dissipating Collisional Clouds of Sub-Micron Grains}

Temporal studies of debris disks offer a unique opportunity to observe the occurrence and dynamical evolution of catastrophic planetesimal collisions in young planetary systems. In order to better understand our sensitivity limits to planetesimal collisional remnants in different regions of the \bpic{} disk over time, we have created a simple planetesimal collisional model to simulate the temporal evolution of the initial sub-micron remnants of planetesimal collisions that can then be injected into our images.

\subsection{Collisional Model}

We use a simple approach to estimate the mass of colliding planetesimals needed to achieve a 5 SNR detection. As this is a complex parameter space, we aim to create a collisional model drawing from more sophisticated modelling approaches \citep[e.g.][]{2020PNAS..117.9712G} using a few simplifying assumptions. Considering these assumptions, we expect our results to be accurate to within an order-of-magnitude.  We begin with the simplifying assumption that the mass of the colliding bodies are the same. We then generate the number of dust particles created from a collision that would catastrophically destroy both colliding bodies in a collisional cascade \citep[e.g.,][]{krivov_cascade}:

\begin{equation}
    \frac{dn(s)}{ds} \propto s^{-3.5}
\end{equation}

where $n$ is the number of particles and $s$ is the radii of detectable dust particles with STIS, which is 0.07-0.7 $\mu$m \citep{2020PNAS..117.9712G}. We use this size distribution to compute how many particles between $s$=0.07-0.7 $\mu$m would be generated from a collisional cascade from given progenitor masses. For simplicity, all particles in this radii range were then assumed to be 0.2 $\mu$m in radii. This is where the peak in dust emission occurs for STIS' wavelength range for a short period of time after collision before being quickly blown out of the system from radiation pressure \citep{2020PNAS..117.9712G}. Simulations \citep{2014A&A...566L...2K, 2016A&A...587A..88T} show that the shape and lower bound of the size distribution of particles produced by high-velocity impacts are sensitive to material properties and impact energy. Given this, we use 0.07 $\mu$m as our lower limit, but we note that the choice of the smallest grain size can slightly vary the mass of the progenitors needed to obtain the same amount of dust. For example, using a minimum grain size of 0.1 $\mu$m from \cite{2020PNAS..117.9712G}, instead of our value of 0.07 $\mu$m, would require the planetesimals to be approximately 1.8 times more massive in order to produce the same amount of dust as the 0.07 $\mu$m mimimum grain size in our model.  The number of particles produced in this range is ultimately dictated by the largest remnant produced. In our case, we use a 10 km body as the largest remnant \citep{debris_max_size}. Additionally, we assume a density of 3.5 g/cm$^3$ for debris material \citep{debris_density}. We assume the collision occurred between parent bodies that are on Keplerian orbits at a stellocentric distance, $r$. Since we assume a catastrophic collision where both bodies are destroyed, impact angle does not play a role in our model. The particles produced in the collision are then given an additional velocity equal to the escape velocity of the parent body in random directions. 

We then calculate a $\beta$ value, the ratio of the force due to radiation pressure and gravity from \cite{Burns_beta_1979}:

\begin{equation}
    \beta = \frac{3LQ_{pr}}{16\pi GMcs\rho}
\end{equation}

where $L$ is the luminosity of the star, $M$ is the mass of the star, $\rho$ is the density of the dust particle, $s$ is the radius of the dust particle, and $Q_{pr}$ is the radiation pressure efficiency coefficient. $Q_{pr}$, which ranges from 0 (perfect transmitter) to 2 (perfect backscatterer), takes into account the amount of momentum transfer from the radiation onto the dust particle \citep{2006A&A...455..509K}. For large dust grains (> 1 $\mu$m), it can be approximated that the particle is a perfect absorber, which would set Q$_{pr}$ to unity. In order to approximate $\beta$ for a sub-micron sized particle, we use stellar parameters of 1.75 M$_{\odot}$ and 9.3 L$_{\odot}$ for \bpic{} \citep{beta_pic_mass, gaia_beta_pic}. Finally, we calculate $\beta$ for small and large Q$_{pr}$ (0.3 and 1.5) to get a range for our $\beta$ value for a 0.2 $\mu$m sized particle. We find a lower bound of 1.3 and an upper bound of 6.5 for $\beta$. For simplicity, we assume the intermediate value of 3.9 for our $\beta$ in the case of 0.2 $\mu$m silicate grains, and note the range as a factor of uncertainty. Additionally, it should be noted that in this model the $\beta$ value does not change our sensitivity to collisions in the disk since we are using a single grain size, it only changes the final location of the cloud after collision. 

The orbital motion of a dust particle under the influence of radiation pressure is determined  using the photogravitational equation for particles affected by radiation pressure \citep{2006A&A...455..509K}:

\begin{equation}
    F_{pg} = -\frac{GMm(1-\beta)}{r^2}.
\end{equation}

Additionally, we ignore the effects of Poynting-Robertson drag due to the shorter collisional timescale's we are dealing with \citep{2006A&A...455..509K}. Then the brightness of the cloud in scattered light was estimated using the following equation: 

\begin{equation}
    B(I_{inc},d_{cloud},\tau) = p(\theta) \cdot I_{inc} \cdot (\pi (\frac{d_{cloud}}{2})^2) \cdot (1-e^{-\tau}) 
\end{equation}

where $p(\theta)$ is the scattering phase function (which is set to $\frac{1}{4\pi}$ because we assume isotropic scattering), $I_{inc}$ is the the amount of incident radiation on the cloud at a given separation, $d_{cloud}$ is the diameter of the dust cloud, and $\tau$ is the optical depth of the dust cloud. It should be noted that  $d_{cloud}$ varies with time, which in turn changes the amount of light incident on the cloud ($I_{inc}$) as well. Given the total number of particles created within our particle size range and the size of the cloud, we can calculate the number density, $N$, of particles in the cloud as a function of cloud size. The optical depth was found using the relation $\tau = N\sigma d_{cloud}$, where $\sigma$ is the cross-section of each dust particle. Finally, the cloud brightness is converted to a flux density after accounting for the wavelength range of STIS and the distance to \bpic{}. 

\subsection{2D Sensitivity Maps for Planetesimal Collisions} 

\begin{figure*}[h]
    \centering
    \includegraphics[width=\textwidth]{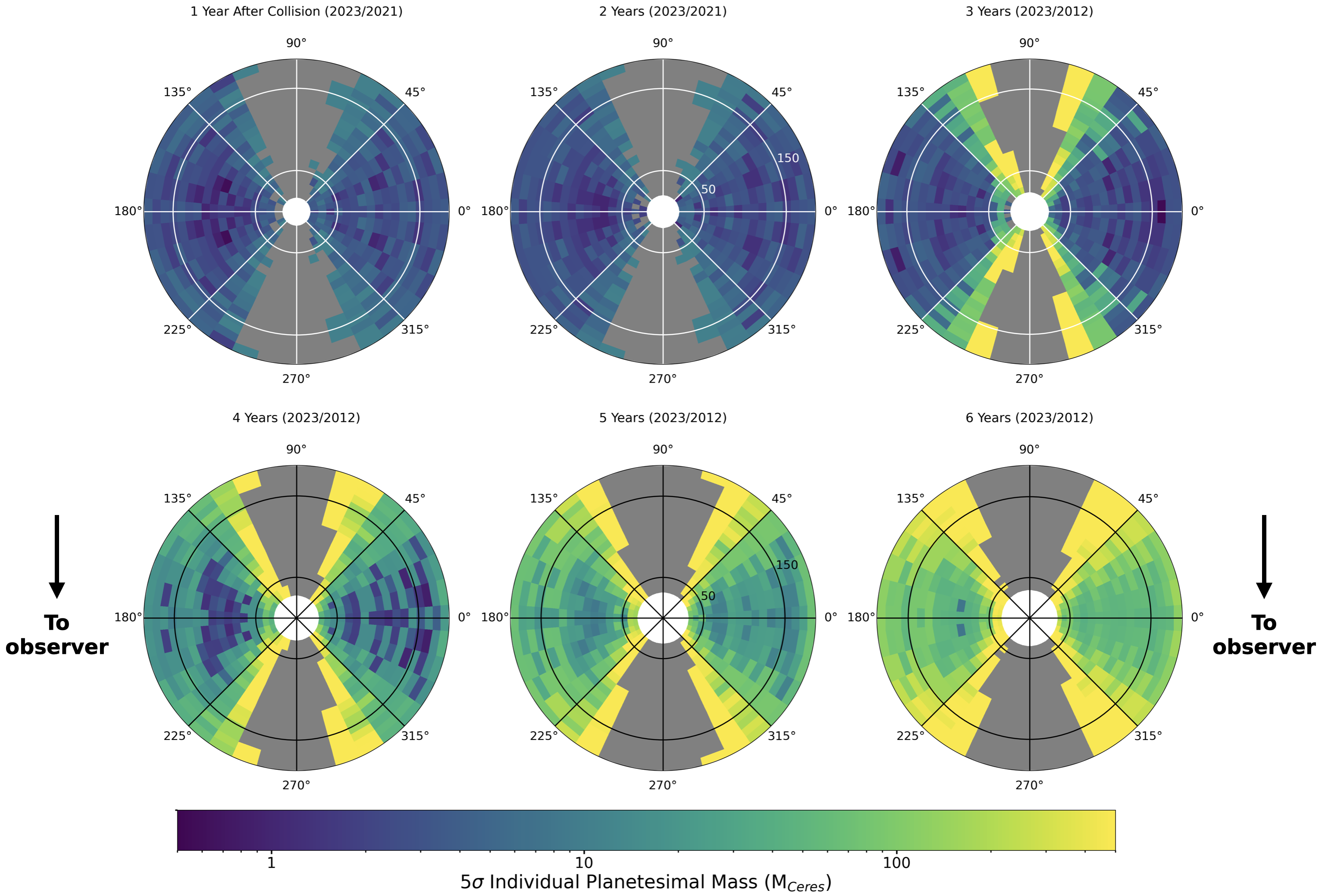}
    \caption{5$\sigma$ 2D collisional sensitivity maps for the \bpic{} disk. Each map shows the mass of a single collisional progenitor needed to have a 5$\sigma$ detection in our ratio maps. The epochs used in the ratio maps is given by the parantheses in each figure title. The grey regions indicate regions in the disk blocked out by our coronagraphic wedge. The radial labels are shown in units of au. The maps are shown from a face-on perspective of the disk, and the black arrows indicate the direction to the observer. We can see a near exponential increase in progenitor mass over the 6 years modeled, pointing to continual monitoring being essential to finding collisional remnants in scattered light observations and observing multi-wavelength evolution of the remnants.}
    \label{fig:contrast_curve}
\end{figure*}

Using the planetesimal collisional model, we are able to understand how sensitive we are to collisions in different regions in the disk. In this case, we want to understand our collisional mass detection limits for different timespans after collisions. For example, if a collision were to have occurred in the \bpic{} disk in 2018, we want to understand how massive the colliding bodies must be in order for us to detect it in different regions of the disk five years after collision in the 2023/2012 ratio map. We do this by evolving collisions occurring at different stellocentric distances (from 20-190 au) over a given number of years and varying progenitor masses. Azimuthal angle in this case refers to the angle that dictates where the collision appears to be in projected separation, where 0$^\circ$ and 180$^\circ$ is when projected and true separation are equal. In this case 180$^\circ$ corresponds to the eastern side of the disk. The brightness and size of the cloud after collision is then convolved with the STIS PSF and injected into the projected location in the disk in the later dataset (2023 for the example above). We take a ratio of the two epochs, in the same way as in Figure \ref{fig:injection}, and then calculate the SNR of the injection. The progenitor mass that has an SNR of 5 in our injections is then chosen as the minimum mass of planetesimal collision we are sensitive to, with the final product seen in Figure \ref{fig:contrast_curve}.

\section{Discussion}

\subsection{Constraining the Azimuthal Structure of the Outer Disk}

Single epoch imaging of debris disks that are edge-on, although easier to resolve due to higher column densities of dust, comes with degeneracies in the azimuthal structure of the disk. Previous work by \cite{2015ApJ...800..136A} measured the surface brightness variations in \bpic{} disk over a period of 15 years, using STIS data from 1997 and the same 2012 data used in our analyses. They showed that a combination of temporal measurements and disk modeling can be used to infer the azimuthal structure of \bpic{}'s edge-on disk. We expand on this work here.

\cite{2015ApJ...800..136A} found the surface brightness of the outer disk midplane, between projected separations of 60-120 au, to vary by $\le$ 3\%. Subsequent modeling of the disk found that small surface brightness changes in the 15 years probed translates to an outer disk that is azimuthally homogenous and lacks isolated structures in scattered light observations. 

Our measurements for surface brightness changes in the \bpic{} disk from 60-120 au tell a very similar story, finding $\le$ 2\% variation from 2012 to 2023. This reaffirms the measurements and modeling done by \cite{2015ApJ...800..136A}, showing the outer disk is largely azimuthally homogeneous, with small percent-level variations. This in turn pushes the baseline of surface brightness measurements in the outer disk of \bpic{} to 26 years.

This is not to say we are not observing real low-level surface brightness variations in the outer disk midplane. We can see in Figure \ref{fig:ratio_grid} that there are regions in the outer disk midplane with surface brightness variations of $\ge 5\sigma$. These variations are most likely due to radiation blowout of small dust grains from other regions of the disk. 

\subsection{Potential Inner Disk Shadowing Below the Western Midplane}

From the 2021/2012 and 2023/2012 surface brightness grid maps, a surface brightness increase was observed below the western midplane of the disk, spanning from 80-200 au, that was not observed on the eastern side of the disk or in the 2023/2021 grid ratio maps.

With our current datasets, we cannot say for certain whether the feature is real or due to image artifacts like a variable stellar halo or PSF subtraction residual. However, if this feature is real, it unlikely that we are observing major dust density changes in the disk due to the orbital period at these separations being $\ge 500$ years. One possibility (assuming an optically thin disk) for the observed brightening is the presence of a low-inclination dust cloud orbiting close-in to \bpic{} at small separations (< 10 au) in the disk, within our inner working angle. Since STIS probes the disk in optical scattered light rather than thermal emission, surface brightness changes are sensitive to the amount of radiation incident on the dust in the disk, in addition to the variations in dust density. A dust cloud passing in front of the western side of the disk in 2012, but not in the 2021 or 2023 observations, could decrease the amount of starlight reaching this region of the disk, explaining the nearly 5\% brightening observed below the western midplane over the last eleven years.  %Future follow-up observations and recovery of the tentatively changing surface brightness .% to ensure that the feature is not to confirm what we are observing is due to flux changes in the disk. 

\subsection{Implications For Planet-Disk Interaction Models }

Current models of planet-disk interactions in debris disks, specifically regarding planetary migration in the presence of a debris disk, have focused on resonance trapping of planetesimals and dust grains by large planets embedded within the disks, showing different induced disk structures for different wavelengths of observations \citep[e.g.,][]{2012A&A...544A..61E}{}{}. Temporal observations of debris disks with STIS, due to its long operational history and stability, offer the best opportunity to observe dust trapped in a resonance with a planet since models show significant dust density changes over time for the particle sizes probed at optical wavelengths. 

Through a wide variety of mechanisms ranging from gas drag to planetesimal scattering, planets can migrate inwards or outwards over time \citep[e.g.][]{migration_review}. While migrating, massive planets are able to capture planetesimals and dust in orbital resonances, causing a clumping of planetesimal distributions in a disk. Through trapping of dust or continual collisions of planetesimals trapped in resonance, these clumpy structures can arise in scattered light observations \citep{2006ApJ...639.1153W,2012A&A...544A..61E}. Once this clumpy dust structure is created, it will trail the planet by approximately 90$^\circ$, and as the planet orbits the star, the structure will change locations, causing major surface brightness variations over time \citep{2006ApJ...639.1153W}. 

In our data, we would expect to see the most drastic surface brightness changes in the separation around \bpic{} b and outward, which is approximately at 10 au (see Figure 3 from \citealt{2006ApJ...639.1153W} to see resonant cloud structure for dust grains with large $\beta$). Furthermore, we would expect to see these changes in the 2021/2012 and 2023/2012 surface brightness comparisons, since that is probing about half \bpic{} b's orbit. This would mean that we could expect the resonant dust structure, if it exists, to change locations from the western side of the disk to the eastern side of the disk. In the 2021/2012 and 2023/2012 grid maps seen in Figure \ref{fig:ratio_grid}, we see surface brightness changes on the order of $6\%$ from 10-30 au in the disk. Due to a combination of photon and PSF subtraction noise, the 3$\sigma$ uncertainty is greater than $6\%$ in this region of the disk.

One solution to the lack of resonant dust cloud is that \bpic{} b underwent little to no migration in the last 12 Myr, meaning the dust cloud never formed and explaining the (lack of) surface brightness changes observed. It could also be that planetesimals trapped in resonance with \bpic{} b are no longer undergoing regular collisions due to a depleted large planetesimal population, making the clump undetectable in scattered light observations. This could be due to the fact that planetesimals in resonant orbits have higher collisional probabilities than those not in resonance \citep{2007CeMDA..99..169Q, 2008A&A...480..551R, 2021MNRAS.503.4767P}, causing massive collisions to be detectable shortly after a potential \bpic{} b migration but not be present today. \cite{2007CeMDA..99..169Q} notes that for the Trojans around Jupiter, the collision rate can be enhanced by an order of magnitude compared to planetesimals not in resonance with a planet. Given \bpic{} b's mass of $12.8^{+5.5}_{-3.2}$ $ M_J$ \citep{2020AJ....159...71N}, the collision rate of planetesimals in resonant orbit with \bpic{} b right after a potential migration could be even higher than that of Jupiter and the Trojans. Another mechanism that could cause a depleted planetesimal population and the non-detection of the resonant dust cloud is that \bpic{} b cleared the dust within its orbital neighborhood very quickly after formation. Modeling by \cite{2015ApJ...811...18M} has shown that \bpic{} b can clear the dust and plantesimals in the disk up to 15 au in only a couple hundred to thousands of orbits (< 0.1 Myr). This would make it such that even if \bpic{} b underwent orbital migration in the last 12 Myr, there would not be sufficient dust to create a detectable resonant dust cloud. 

% see figure from wyatt 2006

\subsection{Sensitivity to Planetesimal Collisions Throughout the \bpic{} Disk Over Time  }

Being such a dynamic disk, \bpic{} serves as a prime target to observe the collisional evolution of planetesimal collisional remnants in real time. In this work, we seek to provide a collisional sensitivity map for planetesimal collisions in the \bpic{} disk in temporal scattered light observations with HST/STIS.

We find that we are sensitive to scattered light from lower mass progenitors in the disk immediately after collision, while more massive planetesimals are needed to detect the remnants many years after collision. More specifically, from our model we find that we are sensitive to Ceres-mass progenitors colliding up to four years after collision in certain regions of the disk (see Figure \ref{fig:contrast_curve}). On the other hand, to detect a collision five years after collision progenitors about an order-of-magnitude more massive than Ceres are required and almost two orders-of-magnitude to detect the remnant after six years. The different detection limits are due to dust production and expansion of the debris cloud post-collision. In the first two years after collision, the dust cloud created is always optically thick for progenitors $> 0.5$ $M_{\text{Ceres}}$. This in turn means the cloud isotropically scatters away almost all incident light in our model. As the cloud expands and remains optically thick, the light incident on the cloud will increase while still scattering most of the light, becoming brighter. However several years after collision, the cloud will have sufficiently expanded and will become optically thin, becoming dimmer per unit area. This dimming depends on the progenitor mass, with more massive collisions remaining brighter for longer.

We find we are most sensitive to the lowest mass collisions in the separations between 50-150 au and at the wings of the disk. This can be most apparent in the fourth panel of Figure \ref{fig:contrast_curve}, where the dark regions indicate sensitivity to lower mass colliding progenitors, and lighter regions indicate sensitivity to more massive colliding progenitors. More massive progenitors are needed in the inner disk due to higher noise levels from PSF subtraction and in the outer disk due to higher levels of photon noise and higher levels of background relative to the scattered light brightness of the collisional remnant. The azimuthal dependence of the progenitor mass comes from projected location of the dust cloud on the disk as opposed to its true separation. For example if a dust cloud was at an orbital separation of 100 au and at an azimuthal angle of 70$^\circ$, it would appear to be at a projected separation of approximately 35 au, which has a background scattered light value (from the overlapping dust along the line of sight) of almost nine times higher than dust at 100 au, which in turn introduces extra noise into the injection measurement.

\subsection{Implications for Collisions in the \bpic{} Disk With Recent JWST Results}

We report in this study that we find no evidence for recent planetesimal collisions in the \bpic{} debris disk, and discuss this within the context of the "Cat's Tail" discovered with JWST/MIRI coronagraphy \citep{JWST_BPic}. The "Cat's Tail" is thought to be the collisional remnants of planetesimals with total masses ranging from 10$^{19}$-10$^{21}$ kg (0.02-2.2 M$_{Ceres}$). While we would be sensitive to upper end of this mass limit in our temporal comparisons 2-3 years after collision (see Figure \ref{fig:contrast_curve}), the authors find the progenitors of the remnants observed with JWST must have collided approximately 150 years ago. Given the fact that small particles, 0.07-0.7 $\mu$m, make up the majority of the flux from dust STIS is sensitive to \citep{2020PNAS..117.9712G}, one potential reason we do not see the collision is because the dust created from a collision 150 years ago in STIS' sensitivity range could be expelled from the system very quickly due to higher radiation pressure experienced by small particles, which in turn would also disperse the cloud below our detection limits more quickly. Further modeling of both datasets together is possible, but beyond the scope of this paper.%However, fully understanding why we do not see the collision seen with MIRI would require in-depth modeling, which is beyond the scope of this paper. 

\subsection{Precision and Future of Time-Differential STIS Coronagraphy}

The coronagraphic observations of the \bpic{} debris disk with STIS in this study has provided the highest precision temporal analyses of a debris disk to date. In the midplane of the disk between 50 and 150 au, we have achieve an SNR of $\ge$ 150. These SNR levels in turn allow us to achieve sub-percent precision in temporal surface brightness changes along the disk midplane from 50-150 au.

The combination of our multi-epoch analysis and collisional modeling seen in Figure \ref{fig:contrast_curve} shows that regular monitoring of dynamic disks like \bpic{} can allow us to discover low-mass planetesimal collisions in the disk in scattered light with high SNR. Such a discovery could then enable follow-up spectroscopy of the system with JWST to better understand the composition of grains and colliding progenitors.

\section{Summary}

In this study, we presented two new (2021 and 2023) epochs of HST/STIS high-contrast coronagraphy of the \bpic{} debris disk and compared the images to a previous STIS dataset obtained through near-identical observations in 2012.
The key results of our study are as follows:

1) Our coronagraphic imaging detects the \bpic{} disk at projected separations between approximately 10-220 au for all three epochs of data. Additionally, we resolve the brightness asymmetry between the east and west sides of the disk that was reported across all wavelengths of observations.

2) We have presented the highest precision and SNR temporal comparison of the \bpic{} debris disk to date, obtaining sub-percent precision for temporal surface brightness variations along large swathes of the disk midplane from 50-150 au in projected separation.

3) We find $\le 2\%$ variation in the disk surface brightness between 3\arcsec{}-6\arcsec{} from 2012 to 2023, reaffirming analysis of the 1997 and 2012 epochs and modeling from \cite{2015ApJ...800..136A} that found only an azimuthally symmetric outer disk can explain such consistent surface brightness measurements from 1997 to 2023.

4) We report the non-detection of a resonant dust cloud trailing \bpic{} b in the inner disk, implying \bpic{} b either did not undergo large scale orbital migration or cleared out the planetesimals near its orbit very early after formation.

5) We report the non-detection of major plantesimal collisions in the disk from 2012 to 2023, while creating a simplified model that generates and evolves the transient sub-micron debris remnants created just after a major planetesimal collision in scattered light.

6) Through injection-recovery tests coupled with simulations of planetesimal collisions, we create 2D collisional sensitivity maps, showing the minimum mass of planetesimal collisions needed to retrieve a 5$\sigma$ detection in the disk ratio images throughout different parts of the disk. We find that the data are sensitive to the least massive planetesimals (approximately a Ceres mass) < 4 years after collision, with the minimum mass of each planetesimal needed for a 5$\sigma$ detection drastically increasing after that. 

7) We discuss these detection limits in the context of other observations of $\beta$ Pic that have suggested major collisions over the past century, and the potential to detect a future collision with continued monitoring.

\section*{Acknowledgements}

We would like to thank Tom Esposito for helpful discussion and guidance on developing our STIS data reduction code. The results reported herein benefited from collaborations and/or information exchange within NASA's Nexus for Exoplanet System Science (NExSS) research coordination network sponsored by NASA's Science Mission Directorate. This research is based on observations made with the NASA/ESA Hubble Space Telescope obtained from the Space Telescope Science Institute, which is operated by the Association of Universities for Research in Astronomy, Inc., under NASA contract NAS 5–26555. These observations are associated with program(s) HST-GO-16174, HST-GO-16788, and HST-GO-16992.

\bibliography{beta_pic}{}
\bibliographystyle{aasjournal}

\appendix

\section{Error Propagation For Extended Sources} 

%Due to the extended nature of resoled debris disks, regular error propagation methods, assuming each pixel is an independent measurement, is not possible. 
Once an error map is generated using the standard error of the mean for each epoch of data, the main operations used in our analysis that requires error propagation is taking a ratio of different epochs and averaging together pixels in an aperture for the radial surface brightness profiles. 

Our error propagation for the two methods described above follows the general error propagation equation for correlated measurements: 

\begin{equation}
    \sigma_{corr}^2 = \sigma_{P_1}^2 (\frac{\partial f}{\partial P_1})^2 +  \sigma_{P_2}^2 (\frac{\partial f}{\partial P_2})^2 + ... + 2 (\frac{\partial f}{\partial P_1}) (\frac{\partial f}{\partial P_2}) \text{Cov}(P_1,P_2) + ...
\end{equation}

where $P_1$ and $P_2$ are individual measurements, $\sigma_{P_1}$ and $\sigma_{P_2}$ are the uncertainties associated with each measurement, $f$ is the functional form of the operation that is applied. 

When applying this to propagating correlated values when taking the mean in an aperture, the functional form becomes $f(P_1,P_2,...) = \frac{P_1 + P_2 + ...}{N}$, where N is the number of pixels in the aperture. The final error propagation equation then becomes: 

\begin{equation}
    \sigma_{corr} = \frac{1}{N} \sqrt{\sigma_{P_1}^2 + \sigma_{P_2}^2 + ... + 2\text{Cov}(P_1,P_2) + ...}
\end{equation}

Additionally when applied to ratio operations ($f(P_1,P_2) = \frac{P_1}{P_2}$), the error propagation then becomes: 

\begin{equation}
    \sigma_{corr} = f(P_1,P_2)  \sqrt{\frac{\sigma_{P_1}^2}{P_1^2} + \frac{\sigma_{P_2}^2}{P_1^2} - \frac{2\text{Cov}(P_1,P_2)}{P_1 P_2}}
\end{equation}

\begin{figure}[h]
    \centering
    \includegraphics[width=\textwidth]{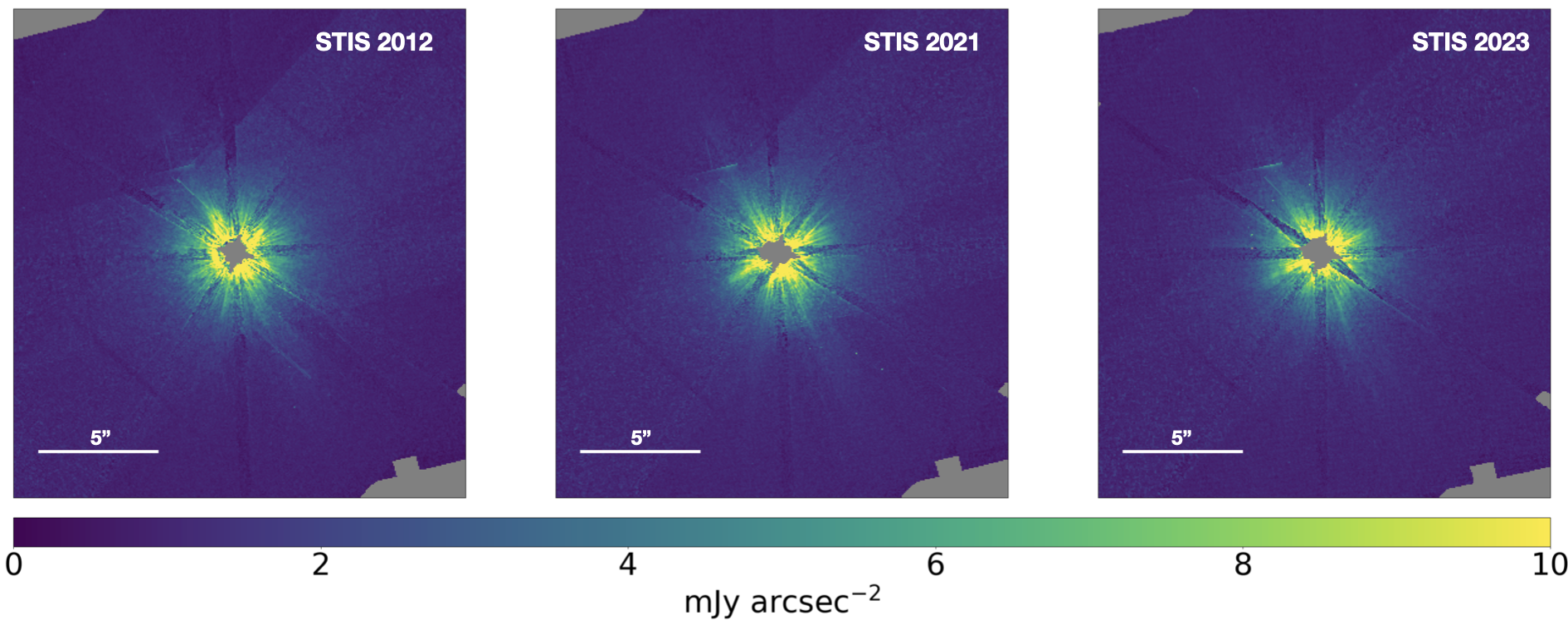}
    \caption{Uncertainty maps for each epoch of data in our sample. The higher levels of noise in the inner disk seen is due to the noise introduced by PSF subtraction residuals and higher levels of photon noise.}
    \label{fig:enter-label}
\end{figure}

%\begin{figure}[!ht]
%    \centering
%    \includegraphics[width=\textwidth]{Figures/error_map_final.png}
%%    \caption{Caption}
 %   \label{fig:enter-label}
%\end{figure}



\end{document}